\documentclass[12pt]{article}
\pdfoutput=1
\usepackage[utf8]{inputenc}
\usepackage{amsmath}
\usepackage{units}
\usepackage{graphicx}
\usepackage{slashed}
\usepackage{array}
\usepackage{amssymb}
\usepackage[leftcaption,raggedright]{sidecap}
\usepackage{caption}
\usepackage[hidelinks]{hyperref}
\usepackage{booktabs}
\usepackage{wrapfig}

\newcommand\pubnumber{NuPhys2026-Will-Parker}
\newcommand\pubdate{\today}

\newcommand{\alphan}{($\alpha, n$) }
\newcommand{\alphap}{$\alpha$-$p$ }

\def\napoli{Department of Physics, University of Oxford, Oxford, United Kingdom}

\def\support{\footnote{
  Science and Technologies Facilities Council.
}}

\def\Title#1{\begin{center} {\Large #1 } \end{center}}
\def\Author#1{\begin{center}{ \sc #1} \end{center}}
\def\Address#1{\begin{center}{ \it #1} \end{center}}

\newcommand\pubblock{\rightline{\begin{tabular}{l} \pubnumber\\
         \pubdate  \end{tabular}}}
\newenvironment{Abstract}{\begin{quotation}  }{\end{quotation}}
\newenvironment{Presented}{\begin{quotation} \begin{center} 
             PRESENTED AT\end{center}\bigskip 
      \begin{center}\begin{large}}{\end{large}\end{center} \end{quotation}}

\def\beq{\begin{equation}}
\def\eeq#1{\label{#1}\end{equation}}
\def\eeqn{\end{equation}}

\def\beqa{\begin{eqnarray}}
\def\eeqa#1{\label{#1}\end{eqnarray}}
\def\eeqan{\end{eqnarray}}

\let\bar=\overbar

\def\Dslash{\not{\hbox{\kern-4pt $D$}}}
\def\dslash{\not{\hbox{\kern-2pt $\del$}}}

\def\msb{{\bar{\ssstyle M \kern -1pt S}}}

\begin{document}
\begin{titlepage}
\pubblock

\vfill
\Title{Reactor Antineutrino Oscillations and Geoneutrinos in SNO+}
\vfill
\Author{William~Parker\support}
\Address{\napoli}
\vfill
\begin{Abstract}
SNO+ is a multipurpose liquid-scintillator neutrino detector located 2~km underground at SNOLAB, Canada. Three large nuclear reactors at baselines of 240–350~km allow a precise measurement of the neutrino oscillation parameter $\Delta m^2_{21}$ and, to a lesser extent, $\theta_{12}$. A spectral analysis is performed, simultaneously fitting $\Delta m^2_{21}$, $\theta_{12}$, the reactor antineutrino flux, background rates, and associated systematics. Using data collected between May 2022 and July 2025, corresponding to a livetime of 685~days, a value of $\Delta m^2_{21} = (7.93^{+0.21}_{-0.24}) \times 10^{-5}$ eV$^2$ is obtained. This result is compatible with other long-baseline reactor antineutrino measurements by KamLAND and JUNO. SNO+ has also made the first measurement of the geoneutrino flux in the Western Hemisphere, measuring $49^{+13}_{-12}$ TNU, in agreement with predictions from geological models.

\end{Abstract}
\vfill
\begin{Presented}
NuPhys2026, Prospects in Neutrino Physics\\
King's College, London, UK,\\ January 7--9, 2026
\end{Presented}
\vfill
\end{titlepage}
\def\thefootnote{\fnsymbol{footnote}}
\setcounter{footnote}{0}

\section{Introduction}

Nuclear reactors produce intense fluxes of electron antineutrinos through the beta decays of fission products. Reactor antineutrino experiments at long baselines provide a precise probe of the solar neutrino oscillation parameters, in particular the mass-squared splitting $\Delta m^2_{21}$ and the mixing angle $\theta_{12}$. The disappearance of electron antineutrinos over distances of order a few hundred kilometres leads to an energy-dependent distortion of the detected inverse beta decay (IBD) signal, allowing these parameters to be extracted through a fit to the reconstructed energy distribution. This technique was first demonstrated by KamLAND~\cite{kamland1} and has since become central to precision measurements of the solar oscillation parameters. SNO+ provides an independent measurement with a distinct reactor baseline distribution and detector systematics, contributing complementary information to the global oscillation picture.

SNO+ is uniquely positioned to perform this measurement. Three large nuclear reactor complexes in Ontario lie at baselines of 240--350~km, contributing approximately 60\% of the total reactor antineutrino flux seen by the experiment.

In addition to reactor antineutrinos, SNO+ is sensitive to geoneutrinos produced by radioactive decays of $^{238}$U and $^{232}$Th within the Earth. These antineutrinos populate the low-energy region of reactor events and are separated statistically in the fit. A measurement of the geoneutrino flux provides information on radiogenic heat in the Earth with a large contribution from the North American continental crust and SNO+ has made the first such measurement in the Western Hemisphere.

This work presents a binned likelihood analysis of reactor antineutrinos and geoneutrinos using data collected between April 2022 and July 2025. The oscillation parameters $\Delta m^2_{21}$ and $\sin^2\theta_{12}$ are extracted simultaneously with background normalisations and systematic uncertainties. The detector, dataset, and analysis procedure are described in the following sections, and the full analysis is described in detail in~\cite{snoplus}.
\section{The SNO+ Detector and Dataset}

SNO+ is located 2~km underground at SNOLAB in Sudbury, Canada, corresponding to an overburden of approximately 6000~m~w.e. The detector consists of a 12~m diameter acrylic vessel (AV) filled with liquid scintillator and viewed by approximately 9300 inward-facing 8-inch photomultiplier tubes (PMTs) mounted on a 17~m diameter geodesic support structure. The surrounding ultra-pure water shielding and laboratory depth provide strong suppression of external and cosmogenic backgrounds. A detailed description of the detector can be found in~\cite{snoplusdet}.

The primary physics goal of SNO+ is a search for neutrinoless double beta decay ($0\nu\beta\beta$) using $^{130}$Te-loaded liquid scintillator. The reactor and geoneutrino analyses presented here are performed during the scintillator phase prior to tellurium loading.

During the period considered in this analysis, the detector operated in two scintillator configurations. In Dataset~I (2022--2023), the scintillator consisted of linear alkylbenzene (LAB) loaded with PPO at a concentration of 2.2~g/L. In Dataset~II (2023--2025), 2.2~mg/L of the wavelength shifter bis-MSB was added to improve light collection and optical stability. The addition of bis-MSB increased the collected photon yield by approximately 50\%, improving the energy resolution. The modified optical response and small changes in background composition require the two datasets to be modelled independently in the spectral analysis.

The depth of SNOLAB reduces the cosmic muon flux to approximately 3 per hour. Residual muons traversing the detector are identified via their large light yield and characteristic timing signatures, and time-based vetoes are applied to suppress cosmogenic backgrounds. 

The results presented here use data collected between May 2022 and July 2025, corresponding to a total livetime of 685~days after data quality selections and veto deadtime (245.8~days in Dataset~I and 439.4~days in Dataset~II).
\section{Signal and Background Model}

Electron antineutrinos in SNO+ are detected via the IBD interaction ($\bar{\nu}_e + p \rightarrow e^+ + n$), which produces a characteristic delayed-coincidence signature. The positron quickly annihilates, forming the prompt scintillation signal. The neutron thermalises and captures predominantly on hydrogen with a mean capture time of approximately 200~$\mu$s, emitting a 2.2~MeV $\gamma$ ray that forms the delayed signal. This prompt--delayed topology provides powerful rejection of uncorrelated backgrounds.

The primary signal arises from reactor antineutrinos. Reactor core thermal powers and operating histories are taken from publicly available IESO~\cite{ieso} and NRC~\cite{nrc} databases, with time-dependent loading factors applied. The emitted antineutrino spectra are constructed isotope-by-isotope, using the Daya Bay--PROSPECT joint fit~\cite{daya_bay_prospect_spectrum} for $^{235}$U and $^{239}$Pu and the Huber--Mueller parameterisation~\cite{huber2011determination,mueller2011improved} for $^{238}$U and $^{241}$Pu. Oscillations are modelled in the three-flavour framework including matter effects for propagation through the Earth's crust, with $\Delta m^2_{21}$ and $\sin^2\theta_{12}$ treated as free parameters in the fit.

Geoneutrinos originating from the decay chains of $^{238}$U and $^{232}$Th within the Earth's interior also contribute to the IBD sample. Their energy distributions overlap with the low-energy region of the reactor signal and are included as separate components in the fit. The $^{238}$U and $^{232}$Th contributions are modelled independently, with a Gaussian constraint applied to their ratio consistent with geophysical expectations~\cite{Wipperfurth:2019idn, Wipperfurth:2018}. The overall geoneutrino normalisations are allowed to vary freely.

Several background components contribute to the selected event sample. The dominant correlated background arises from \alphan reactions on $^{13}$C, in which $\alpha$ particles from $^{210}$Po decays produce neutrons that mimic the IBD delayed-coincidence signature; the rate is determined using a Daya Bay–based neutron yield~\cite{alphanDayaBay} rescaled to SNO+ and the measured $^{210}$Po activity. An \alphan classifier exploiting event topology and timing information is applied to further suppress this background, and fits are performed both with and without this requirement.

A smaller correlated contribution comes from \alphap events, a background for IBD measurements in scintillator identified by SNO+~\cite{sno+ppo}. Atmospheric neutrino interactions can also produce neutron-emitting events that satisfy the IBD selection and are modelled using simulation. Accidental coincidences from uncorrelated events are estimated using a data-driven approach and included as a background component in the fit.

A binned likelihood fit to the reconstructed prompt energy distribution is performed simultaneously for both datasets, including contributions from reactor antineutrinos, geoneutrinos, and all relevant background components.
\section{Event Selection}

IBD candidates are identified by requiring pairs of reconstructed events consistent with the prompt--delayed topology described in Section~3. Both prompt and delayed events are required to have valid vertex fits and to pass standard data cleaning criteria to remove instrumental backgrounds.

Selection requirements are applied to reconstructed energy, vertex position, spatial separation ($\Delta R$), and time separation ($\Delta t$) between the prompt and delayed signals. Prompt and delayed energy windows are chosen to encompass the expected IBD signal region while rejecting low-energy radioactivity and high-energy cosmogenic backgrounds. A fiducial volume requirement of $R < 5.7$~m is imposed on both events to reduce backgrounds originating near the acrylic vessel and surrounding detector materials.

Coincidence requirements of $\Delta t < 2$~ms and $\Delta R < 2.5$~m suppress accidental coincidences while maintaining high signal efficiency.

Several vetoes are applied to mitigate cosmogenic backgrounds. Events are rejected within 20~s of identified muons to suppress long-lived spallation isotopes. Additional shorter veto windows, including a veto following events with a high number of hit PMTs and a dedicated muon spallation cut, target fast neutron and instrumental backgrounds. A multiplicity requirement ensures no additional activity occurs near the candidate pair.

Residual backgrounds from $^{214}$Bi--$^{214}$Po coincidences are suppressed using dedicated tagging criteria. Further discrimination against accidental and \alphan backgrounds is achieved using likelihood-based and multivariate classifiers. The optimal thresholds for these classifiers differ between the PPO and bis-MSB datasets due to changes in light yield and detector response. The timing-based \alphan classifier is a novel technique applied for the first time in this analysis.
\section{Systematic Uncertainties}

Systematic uncertainties are incorporated in the fit as Gaussian-constrained nuisance parameters. The dominant rate uncertainty on the reactor signal arises from the overall flux normalisation, while the geoneutrino $^{238}$U and $^{232}$Th components are constrained through a prior on their ratio. The \alphan and other background normalisations are included with conservative constraints derived from simulation and data-driven studies.

Detector-related systematics include uncertainties on the IBD selection efficiency and the reconstructed energy scale and resolution, which affect both the overall rate and the shape of the energy distribution.  All systematic parameters are floated simultaneously with the oscillation parameters in the likelihood fit.
\section{Results}

A binned Poisson log-likelihood is constructed from the reconstructed prompt energy distribution and evaluated simultaneously for both datasets. The oscillation parameters $\Delta m^2_{21}$ and $\sin^2\theta_{12}$ are fitted together with signal normalisations, background rates, and systematic nuisance parameters. The best-fit values and confidence intervals are obtained from the likelihood surface.

Fits are performed in three configurations: without oscillation priors, with PDG~\cite{PDG2025} priors on $\Delta m^2_{21}$ and $\sin^2\theta_{12}$, and with PDG priors together with the application of the \alphan classifier.

\begin{figure}[!htbp]
    \centering
    \caption{Best-fit energy distribution for the fit with PDG constraints on oscillation parameters and using the \alphan classifier, which is applied only below 3.5~MeV (dashed line). The reactor IBD distribution is split by reactor complex baseline.}
    \includegraphics[width=0.6\linewidth]{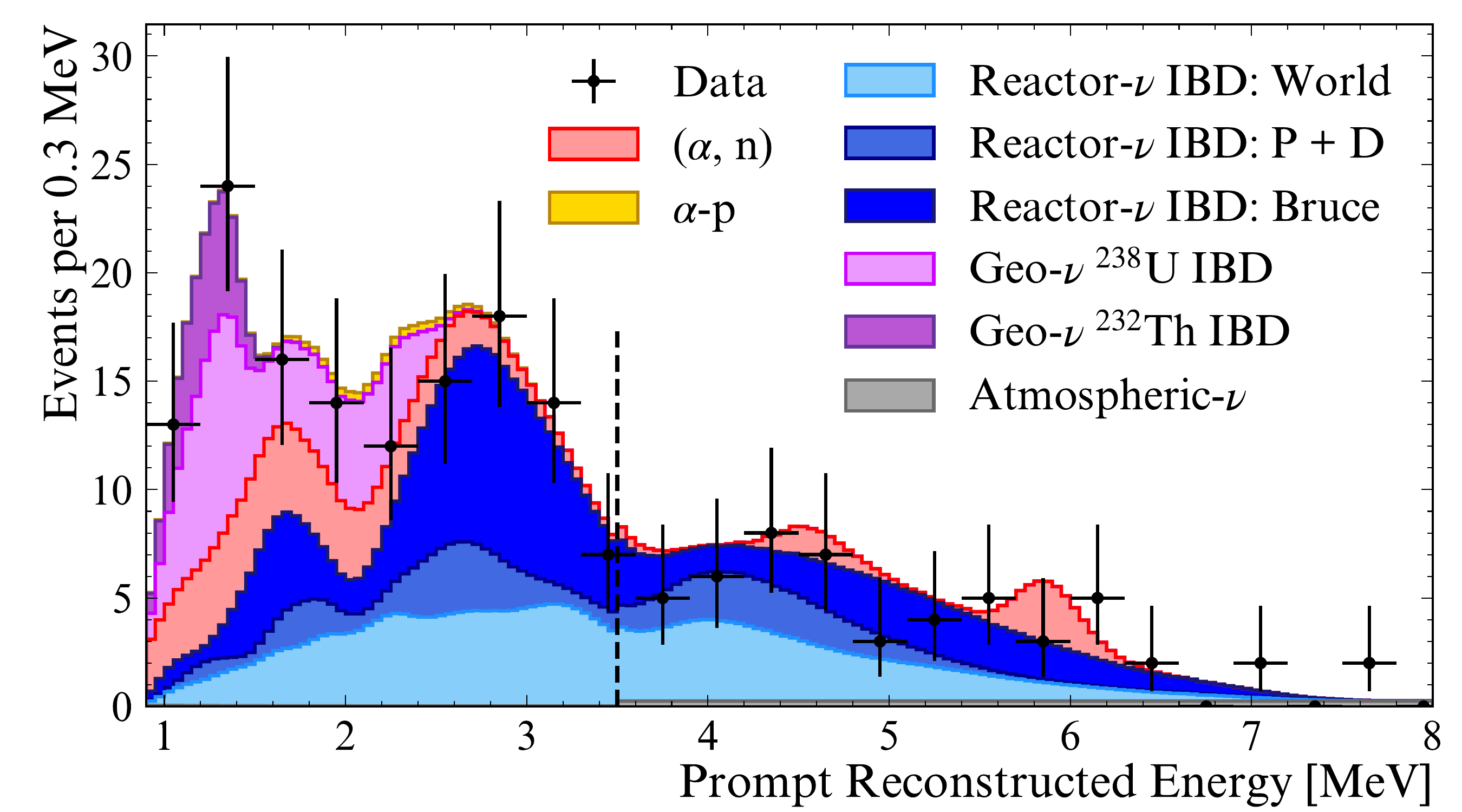}
    \label{fig:Eprompt}
\end{figure}

The reconstructed prompt energy distribution for both datasets is shown in Figure~\ref{fig:Eprompt}, overlaid with the best-fit model. The contributions from reactor antineutrinos, geoneutrinos, and background components are shown separately. Good agreement is observed across the full analysis range.

\begin{figure}
    \centering
    \caption{2-D $\log$-likelihood contours of $\Delta m^2_{21}$ vs. $\sin^2\theta_{12}$, with 1-D $\log$-likelihood profiles. Results are shown for this analysis, as well as results from KamLAND~\cite{kamland_on_off} and combined solar measurements~\cite{Super-Kamiokande_solar}, along with the combined fit of SNO+ and the PDG 2025 global results~\cite{PDG2025}.}
    \includegraphics[width=0.6\linewidth]{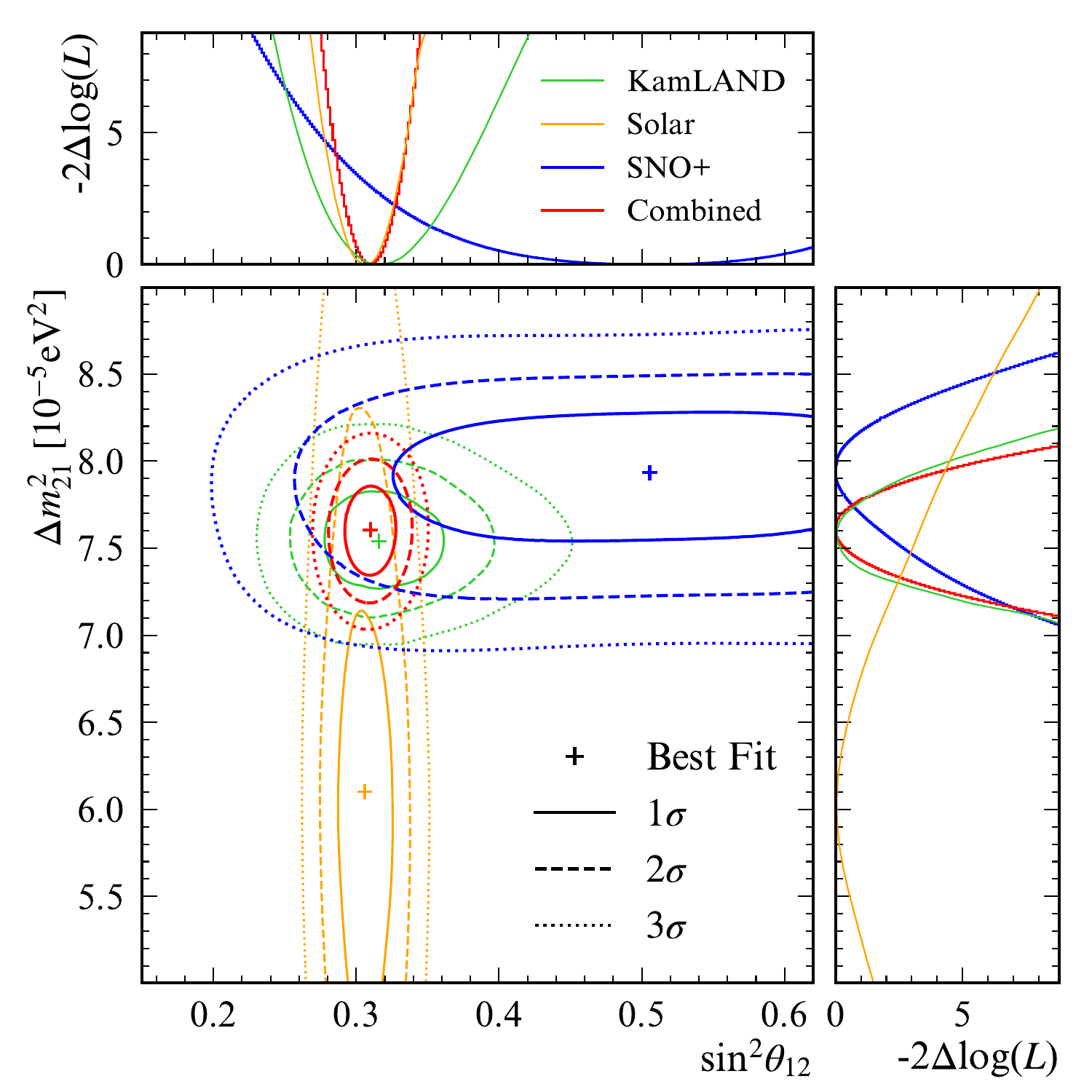}
    \label{fig:contours}
\end{figure}

The two-dimensional likelihood contours in the oscillation parameters are shown in Figure~\ref{fig:contours}, along with results from KamLAND~\cite{kamland_on_off} and combined solar measurements~\cite{Super-Kamiokande_solar}, as well as the combined fit of SNO+ and the PDG 2025 global results~\cite{PDG2025}. The best-fit values are 
$\Delta m^2_{21} = (7.60\pm0.17) \times 10^{-5}~\mathrm{eV}^2$, and $\sin^2\theta_{12}=0.310\pm0.012$. The numerical results are summarised in Table~\ref{tab:fit_results}. The measurement is consistent with other long-baseline reactor experiments.

\begin{table}
\small
\centering
\caption{Expected and fitted event rates and oscillation parameters for fits using SNO+ only, PDG priors on oscillation parameters, and oscillation priors with using the \alphan classifier.} \label{tab:fit_results}
\begin{tabular}{lcccc}
\toprule
\toprule
   & Expected & Fit & Fit (con.) & Fit \alphan cut \\
\toprule
Reactor IBD & $140\pm6$ & $120\pm5$ & $136\pm6$ & $130_{-5}^{+6}$ \\
Geo $^{238}$U IBD & 29 & $38_{-14}^{+15}$ & $38_{-14}^{+15}$  & $27_{-7}^{+8}$ \\
Geo $^{232}$Th IBD & 8 & $11\pm5$  & $12_{-5}^{+6}$ & $8_{-3}^{+2}$ \\
\alphan $p$-scatters & $80 \pm 24$ & $63\pm19$ & $58_{-19}^{+20}$ & $22\pm6$ \\
\alphan other & $12 \pm 10$ & $7\pm5$ & $7\pm5$ & $7\pm4$ \\
\alphap  & $7\pm 6$ & $4_{-4}^{+6}$ & $3_{-3}^{+6}$ & $3_{-3}^{+6}$ \\
Atmospheric $\nu$ & $4\pm 3$ & $6_{-2}^{+3}$ & $6\pm3$ & $5\pm2$ \\
\midrule
Sum & 281 & 249 & 259 & 202 \\
\midrule
Observed  & 246 & 246 & 246 & 185 \\
\midrule
$\Delta m^2_{21}$ [$\times 10^{-5} \text{eV}^2$] & $7.50\pm0.19$ & $7.93^{+0.21}_{-0.24}$ & $7.60\pm0.17$ & $7.56\pm0.17$ \\
$\sin^2\theta_{12}$ & $0.307\pm0.012$ & 0.505$\pm$0.134 & $0.310\pm0.012$ & $0.311\pm0.012$ \\
Geo-$\overline{\nu}$ [TNU] & $46$ & $60^{+23}_{-22}$  & $61^{+23}_{-22}$ & $49^{+13}_{-12}$ \\
Geo-$\overline{\nu}$ U/Th & $3.7\pm1.3$ & $3.38_{-1.41}^{+1.39}$ & $3.30_{-1.44}^{+1.41}$ & $3.29_{-1.48}^{+1.42}$ \\
\bottomrule
\bottomrule
\end{tabular}
\end{table}

The fitted geoneutrino flux corresponds to a total rate of 49$^{+13}_{-12}$~TNU, with individual $^{238}$U and $^{232}$Th components determined under the imposed ratio constraint. The results are summarised in Table~\ref{tab:fit_results} and are consistent with geological model predictions.
\section{Conclusions and Outlook}

A measurement of reactor antineutrino oscillations and geoneutrinos has been performed using 685 days of SNO+ scintillator-phase data. The oscillation parameters $\Delta m^2_{21}$ and $\sin^2\theta_{12}$ are determined from a global likelihood fit to the prompt energy distribution, with results consistent with previous long-baseline reactor experiments.

The fitted geoneutrino flux is consistent with geological model predictions, and SNO+ is the only experiment measuring geoneutrinos in the Western Hemisphere.

The precision of solar oscillation parameters is being further improved by the JUNO experiment~\cite{JUNO:2025gmd}, which operates with a larger detector and shorter reactor baselines. The SNO+ measurement remains complementary, probing antineutrino oscillations over longer distances and with different dominant backgrounds and systematics. Continued data-taking and improved modelling of \alphan and \alphap interactions will further strengthen future SNO+ results.

\bibliographystyle{unsrt} 
\bibliography{ref}

\end{document}